\documentclass[conference]{IEEEtran}
\usepackage{silence}
\usepackage{amsmath, amssymb}
\usepackage{amsthm} 
\usepackage{graphicx}
\usepackage{hyperref}
\usepackage{xcolor}
\usepackage{cite}
\usepackage{float}
\usepackage{comment}
\usepackage[letterpaper,top=0.70in,left=0.667in,right=0.673in,bottom=1.00in]{geometry}
\usepackage{balance}
\usepackage{cite}
\usepackage{amsmath,amssymb,amsfonts}
\usepackage[ruled,vlined]{algorithm2e}
\usepackage{booktabs}   
\usepackage{caption}    
\usepackage{tabularx}   

\def\s{{\mathbf s}}
\def\z{{\mathbf z}}
\def\n{{\mathbf n}}
\def\v{{\mathbf v}}
\def\x{{\mathbf x}}
\def\y{{\mathbf y}}
\def\X{{\mathbf X}}
\def\Y{{\mathbf Y}}
\def\R{{\mathbf R}}
\def\P{{\mathbf P}}
\def\F{{\mathbf F}}
\def\G{{\mathbf G}}
\def\Q{{\mathbf Q}}
\def\H{{\mathbf H}}

\def\U{{\mathbf U}}
\def\V{{\mathbf V}}
\def\I{{\mathbf I}}

\def\S{{\mathbf S}}

\newcommand\independent{\protect\mathpalette{\protect\independenT}{\perp}}
\def\independenT#1#2{\mathrel{\rlap{$#1#2$}\mkern2mu{#1#2}}}

\title{Distributed Semantic Alignment over Interference Channels: A Game-Theoretic Approach}
\author{Giuseppe Di Poce$^{1}$, Mattia Merluzzi$^{1}$, Emilio Calvanese Strinati$^1$, and Paolo Di Lorenzo$^{2,3}$ \medskip \\
$^1$ CEA Leti, University Grenoble Alpes, 38000, Grenoble, France.\\
$^2$ Consorzio Nazionale Interuniversitario per le Telecomunicazioni (CNIT), Parma, Italy.\\
$^3$ DIET Department, Sapienza University of Rome, Via Eudossiana 18, Rome, Italy. \smallskip\\
e-mail: \{giuseppe.dipoce, mattia.merluzzi, emilio.calvanese-strinati\}@cea.fr, paolo.dilorenzo@uniroma1.it.
\thanks{\small This work was funded by SNS JU project 6G-GOALS-EU’s Horizon program (No.101139232), NextGenerationEU RESTART (PE00000001), 
SNS JU 6GARROW (Horizon, n. 101192194), 
French government (2030 ANR, ref. 22-PEFT-0010).}
}

\begin{document}
\IEEEoverridecommandlockouts
\maketitle
\begin{abstract}
    Semantic communication acts as a key enabler for effective task execution in AI-driven systems,
    prioritizing the extraction of the underlying meaning before transmission.
    However, when devices rely on different logic and internal representations, semantic mismatches may arise, potentially hindering mutual understanding and effectiveness of communication.
    Furthermore,  in interference channel environments,
    the coexistence of multiple devices 
    introduce a significant degradation due to the presence of multi-user-interference.
    To address these challenges, in this paper we formulate the joint optimization of linear Multiple-Input-Multiple-Output (MIMO) transceivers as a distributed non-cooperative game, enabling a closed-form solution that effectively addresses semantic coexistence and latent space misalignment.
    We derive sufficient conditions for the existence of a Nash Equilibrium (NE), considering multiple 
    point-to-point MIMO channels, with corresponding users modeled as selfish players optimizing their 
    transmission and semantic alignment strategies. Numerical results substantiate the proposed approach in goal-oriented semantic communication by highlighting crucial trade-offs between information compression, interference mitigation, semantic alignment, and task performance.
\end{abstract}
\smallskip

\IEEEpeerreviewmaketitle
\section{Introduction}
Digital communication systems have been designed to transmit symbols over noisy channels, with the goal of retrieving the exact information at the receiver. 
Information theory, grounded in Shannon's theorems, establishes the fundamental limits on data rate and compression for reliable point-to-point communication systems, forming the core engineering challenge in their design. Semantic and goal-oriented communications (SC) act beyond the mere physical communication problem, focusing on the effective mutual understanding and task execution \cite{gunduz2022beyond}. This paradigm is particularly relevant for AI-native systems, in which different type of intelligent agents interact to reach (possibly common) objectives. In this goal-oriented scenario, future 6G networks employ deep neural networks (DNNs) as crucial feature extractors to represent raw data into a lower-dimensional, semantically meaningful, task-relevant latent space.
Semantic communication therefore offers a promising approach to overcoming bandwidth limitations and resource saturation in wireless systems, which must handle rapidly growing data traffic while supporting stringent latency requirements. Recent research exploits generative-AI\cite{barbarossa2023semantic}, deep joint source channel coding \cite{bourtsoulatze2019deep}, goal-oriented data compression\cite{wanasekara2025sc}, and semantic network coverage \cite{merluzzi2025semantic} to tight network design to AI-native applications.

\noindent Introducing semantics into communication systems creates new challenges, particularly \textit{semantic noise}—errors caused by mismatched logic, interpretation, or background knowledge among AI agents. Specifically, AI-native communication requires the transmitter and receiver to share compatible latent representations. Yet, differences in model architectures or training often lead to divergent representations of the same information, undermining mutual understanding. This issue is especially prevalent in multi-vendor environments where sharing models or training data is impractical or restricted.
Semantic channel equalization \cite{sana2023semantic} 
addresses this problem, aligning latent representations of encoder and decoder when they are not jointly trained, exploiting relative representations~\cite{moschella2022relative,fiorellino2024dynamic},
Parseval frames\cite{fiorellino2025frame}, and Optimal-Transport theory\cite{alvarez2019towards}. This equalization problem has also gained attention to the advent of smart devices, with their own local AI-capabilities stored in silicon, such as in \textit{NeuroCorgi} \cite{panades2024772muj}. These neuromorphic computing devices offer dramatic energy savings to AI-driven task execution and feature extractions.
However, the persistent problem is how to equalize the semantic mismatch generated by fixed devices' DNNs backbone, which cannot be fine-tuned to adapt to different logic.
In this context, proposed solutions should moreover consider the overlap of received signals in space, time and frequency \cite{merluzzi2025goal} due to the coexistence of multiple sources, transmitting to as many destinations over interference channels. In traditional semantic-agnostic systems, this problem has been widely investigated, casting the maximization of mutual information as a non-cooperative game among (cognitive) radio nodes\cite{scutari2008competitive,pang2010design}, competing to optimize their utility function. In the presence of both semantic and channel noise, interference can further degrade performance, leading to catastrophic target-
task execution.

\noindent \textbf{Contributions.}
This work proposes a novel method for multi-user semantic channel equalization over interference channels, by jointly optimizing MIMO semantic linear transceivers. Specifically, under the interference perceived by other users, we formulate a convex optimization problem at each agent whose closed-form solution yields latent-space alignment and power allocation that explicitly accounts for cross-user coupling in the semantic domain.
The coexistence of multiple transmitters and receivers in an interference environment is then cast as a non-cooperative game, in which each player selfishly optimizes its alignment strategy to reach a NE. This approach promotes interference mitigation, implements semantic alignment and compression, thus enabling task-oriented communication under power and complexity constraints. Numerical results validate the proposed approach, revealing key trade-offs among information compression, interference mitigation, semantic alignment, and task performance. To the best of our knowledge, this is the first work addressing distributed semantic equalization over interference channels.

\section{System Model}
\begin{figure}[t]
  \centering
\includegraphics[width=\columnwidth]{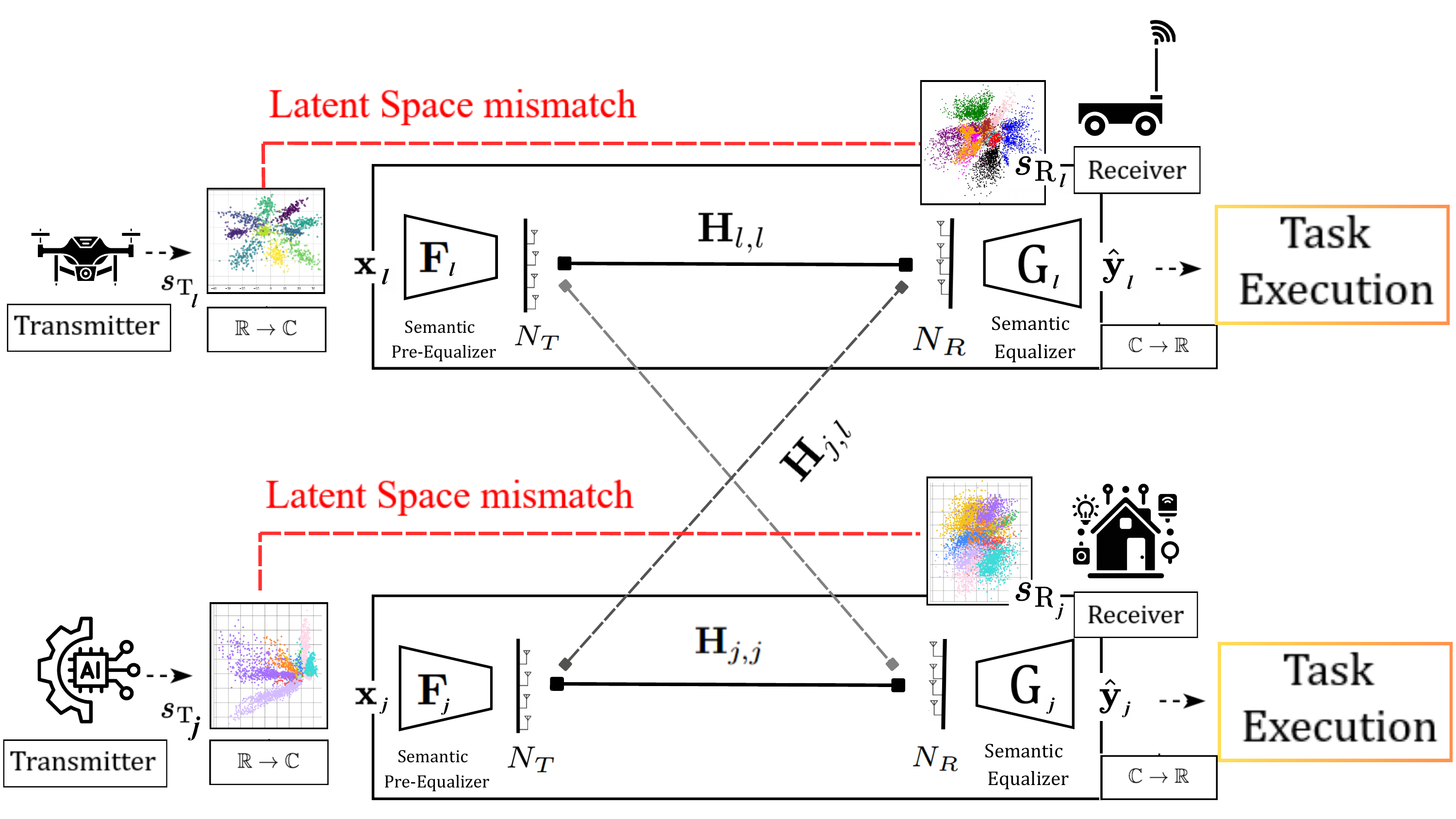}
  \caption{\small Overview of the proposed goal-oriented system model.}
  \label{fig:System_Model}
\end{figure}

We consider the interference channel system depicted in Fig.\ref{fig:System_Model}, which consists of $L$ transmitter-receiver (tx-rx) pairs, indexed by $l$.
Each tx-rx pair is endowed with pre-trained DNNs to encode and decode semantic information, respectively. 
Let $\mathcal{D}$ be a shared dataset accessible to both transmitter and receiver, and $\s_{T_{l}} \in \mathbb{R}^{d_l}$ denote the semantic feature vector extracted by the $l$-th transmitter from a data point $\z \in \mathbb{R}^q$, via a DNN backbone function $\omega_{T_l}(\cdot)$. 
The set of all latent vectors $\s_{T_{l}}$ associated to every $\z\in\mathcal{D}$ represents the transmitter semantic latent space, which encodes its internal logic in order to map the raw input data into a lower-dimensional, task-oriented representation. Each $l$-th receiver relies on its own latent space structure, which differs from that of its intended transmitter and must therefore be properly aligned with the transmitter’s latent space to ensure reliable task performance. Furthermore, simultaneous transmissions introduce multi-user interference (MUI), which degrades the quality of the transmitted latent information. Therefore, the objective is to maximize the alignment between the latent spaces of each tx-rx pair, while accounting for both noise and MUI. 
The proposed semantic equalization scheme at the $l$-th transmitter is composed of the following steps. First, assuming w.l.o.g. that $d_l$ is even, we proceed by pairing the first half of the semantic features in $\mathbf{s}_{T_l}\!\in \!\mathbb{R}^{d_l}$ with the second half to form complex symbols, yielding an input vector $\x_l \!\in\! \mathbb{C}^\frac{d_l}{2}$. Then, we exploit a semantic pre-equalizer $f_l(\cdot)$ that jointly performs semantic alignment and feature compression. Specifically, $f_l\!:\! \mathbb{C}^{\frac{d_l}{2}} \!\xrightarrow{}\! \mathbb{C}^{KN_{T_l}}$ performs a learnable transformation that maps the transmitter complex vector $\x_l\!\in\!\mathbb{C}^\frac{d_l}{2}$ into the compressed representation $\overline{\mathbf{x}}_l\in\mathbb{C}^{K N_{T_l}}$, where $N_{T_l}$ is the number of antennas at the $l$-th transmitter, and $K$ represents the number of channel uses. The compression factor resulting from transmitting $K$ MIMO symbols instead of $d_l/2$ complex values is given by
$\xi_l \!=\! \frac{K}{d_l/2}$. The compressed vectors $\overline{\mathbf{x}}_l$ are transmitted over $K$ channel uses through a flat-fading MIMO channel $\overline{\mathbf{H}}_{l,l} \in \mathbb{C}^{N_{R_l} \times N_{T_l}}$, where $N_{R_l}$ is the number of antennas at the $l$-th receiver. In addition, this transmission causes interference to the other users through the cross-link MIMO channels $\mathbf{H}_{l,j}$, for all $j \!\neq\! l$. Finally, at the rx side, a post-equalization function maps the received symbols into a complex vector $\hat{\y_l} \!\in\! \mathbb{C}^{\frac{m_l}{2}}$ via learnable transformation $ g_l \!:\! \mathbb{C}^{KN_{R_l}} \!\xrightarrow{}\! \mathbb{C}^{\frac{m_l}{2}}$.  Overall, the semantic MIMO communication over the $l$-th tx-rx pair can be modeled as:
\begin{equation}
  \hat{\mathbf y}_l
  = g_l\!\left(
      \mathbf H_{l,l} f_l(\mathbf x_l)
      + \sum\nolimits_{j \ne l} \mathbf H_{j,l} f_j(\mathbf x_j)
      + \mathbf v_l
    \right)
    \label{eq:received_signal_vectorized}
\end{equation} 
where $\H_{l,l} \!=\! \I_K  \otimes \overline{\H}_{l,l} \in \mathbb{C}^{KN_{R_l} \times KN_{T_l}}$ is the direct channel of link $l$, $\H_{j,l} = \I_K  \otimes \overline{\H}_{j,l}  \in \mathbb{C}^{KN_{R_j} \times KN_{T_l}}$ is the cross-channel matrix between source $j$ and destination $l$,  $\v_l$ is a zero-mean circularly symmetric complex Gaussian noise vector with covariance matrix $\R_v\!=\!\sigma^2 \I$.
The second term on the right-hand side of $(\ref{eq:received_signal_vectorized})$ represents the MUI received by the $l$-th destination and caused by the other active communication links. To ease notation, in the sequel we denote the MUI plus noise (MUIN) term of the $l$-th link in $(\ref{eq:received_signal_vectorized})$ as $\n_l= \sum_{j \ne l} \mathbf H_{j,l} f_j(\mathbf x_j) + \mathbf v_l $.
At the $l$-th receiver, the complex latent vector  $\hat{\y_l} \in \mathbb{C}^{\frac{m_l}{2}}$ in $(\ref{eq:received_signal_vectorized})$ is then converted into a real vector $\hat{\s}_{R_l}$ of dimension $m_l$, by inverting the halving operation done at transmitter side. Finally, the received signal is processed by a task-specific neural network at the receiver, which is trained end-to-end to accomplish the desired downstream objective.

Our aim is to jointly optimize the learnable semantic transformations $f_l$ and $g_l$ for each user $l$, aiming to minimize the semantic discrepancy between the true latent vectors $\mathbf{s}_{R_l}$ and the received vectors $\hat{\mathbf{s}}_{R_l}$ across all $L$ communication links. To this end, we exploit latent training vectors as \textit{semantic pilots} to enable logic--channel estimation and alignment. For the $l$-th link, we consider $n$ labeled examples 
\(
\{ (\mathbf{x}_{i,l}, \mathbf{y}_{i,l}) \}_{i \in \mathcal{T}_r}
\)
extracted from the available training set $\mathcal{T}_r$. In the following, we provide a detailed problem formulation that enables a closed-form solution for semantic channel equalization, explicitly taking into account MUI mitigation.

\section{Game-theoretic Semantic Equalization} 

We consider linear semantic pre-equalizers $\{f_l(\cdot)\}_{l=1}^L$, each represented by a matrix $\mathbf{F}_l \in \mathbb{C}^{K N_{T_l} \times \frac{d_l}{2}}$, and linear semantic equalizers $\{g_l(\cdot)\}_{l=1}^L$, modeled by matrices $\mathbf{G}_l \in \mathbb{C}^{\frac{m_l}{2} \times K N_{R_l}}$, for all $l = 1, \ldots, L$. For each $l$-th transmitter, the total average transmit power is constrained as:
\begin{equation}
 \mathbb{E}\{ ||\F_l\x_l||_F^2 \}= \text{tr} \{\F_l\F_l^H \} \leq KP_{\max},\quad \forall l \in L  
    \label{eq:power_constraint}
\end{equation}
where $P_{\max}$ is the power budget per channel use.
We make the following reasonable assumptions:
\begin{itemize}
    \item[A0)] Latent space vectors $\x_l,\y_l$ are modeled as zero-mean random vectors with covariance $\R_x,\R_y = \I$. There is no loss of generality here, as we could always incorporate a whitening 
    matrix as a right factor of $\F_l$;  
    \label{assumption 0: pre-withening}
    \item[A1)] Independence between latent representation transmitted by different communication links:
\begin{equation}
\{\x_l,\y_l\} \independent \x_j \qquad \forall\, l \neq j \in L.
\label{assumption:independence}
\end{equation}
     \item[A2)] We assume perfect channel state information (CSI),  with a sufficiently long channel coherence time such that each channel realization can be regarded as a fixed snapshot;
    \label{assumption 1: stat. independence}
\end{itemize}
Under these assumptions, the channel model over the $l$-th communication link $(\ref{eq:received_signal_vectorized})$ boils down to:     
\begin{equation}
  \hat{\y}_l
    = \G_l\H_{l,l} \F_l \x_l + \G_l\n_l,
    \quad l=1,\ldots,L,
    \label{eq:received_signal_matrices}
\end{equation}
where, due to A1, the covariance of the MUIN term reads as:
\begin{equation}
    \R_{n_l} \triangleq \sum\nolimits_{j \neq l} (\H_{j,l} \F_j)(\H_{j,l} \F_j)^H + \sigma^2 \I.\label{eq:MUI_covariance}
\end{equation}
Once the MUIN term $\mathbf{n}_l$ in \eqref{eq:received_signal_matrices} is locally estimated at user $l$, the design of the semantic transceivers $\mathbf{F}_l$ and $\mathbf{G}_l$ can be formulated as a mean-squared error minimization over the semantic pilot set, under transmit power constraints: 
\begin{align}
\min_{\mathbf{F}_l, \mathbf{G}_l} \;\;
& \frac{1}{n} 
\sum_{i \in \mathcal{T}_r} \mathbb{E}
\left\|
\mathbf{y}_{i,l}
-
\mathbf{G}_l
\left(
\mathbf{H}_{l,l} \mathbf{F}_l \mathbf{x}_{i,l}
+
\mathbf{n}_l
\right)
\right\|_F^2
\notag \\
&\text{s.t.} \quad
 \mathrm{tr}\!\left\{\mathbf{F}_l \mathbf{F}_l^H\right\}
\leq K P_{\max}, 
\label{eq: non-convex ERM}
\end{align}
for all $l = 1,\ldots,L$. The formulation in (\ref{eq: non-convex ERM}) results in a non-convex optimization problem, since the Frobenius-norm objective depends bilinearly on both the pre-equalizer and equalizer matrices. In the following, we develop a solution strategy that convexifies problem~\eqref{eq: non-convex ERM}, enabling the joint optimization of the semantic MIMO linear transceivers.



\subsection{
Optimal Semantic Pre-equalizer}
We start observing that, for fixed $\mathbf{F}_l$, problem~\eqref{eq: non-convex ERM} becomes convex in $\mathbf{G}_l$. Let $\X_l\in\mathbb{C}^{\frac{d_l}{2}\times n}$ denote the matrix that collects all latent transmitter samples $\{\mathbf{x}_{i,l}\}_{i\in\mathcal{T}_r}$,
and $\Y_l\in\mathbb{C}^{\frac{m_l}{2}\times n}$ be the matrix containing the corresponding receiver latent column vectors $\{\mathbf{y}_{i,l}\}_{i\in\mathcal{T}_r}$. Then, the optimal equalizer admits a closed-form solution given by the Wiener filter:
\begin{align}
    \G_l^{opt} = \Y_l \X_l ^H (\H_{l,l}\F_l)^H  (\H_{l,l}\F_l 
    \F_l^H \H_{l,l}^H + n \R_{n_l})^{-1},
    \label{eq:Wiener_filter}
\end{align}
which is optimal at the $l$-th receiver for any given semantic pre-equalizer employed at transmission. Leveraging the closed-form expression in \eqref{eq:Wiener_filter}, and defining the semantic cross-covariance matrix $\P_l = \Y_l \X_l^H $ the $l$-th objective term in~\eqref{eq: non-convex ERM} can be rewritten — after standard algebraic manipulations, omitted due to lack of space — as a function of the pre-equalizer $\mathbf{F}_l$ only. Specifically, we get:
\begin{equation}
    \text{MSE}_l(\F_l)\! = \! \frac{1}{n} \text{tr} \{ \R_{y_l}\! -\!  \P_l \P_l^H  +  (\F_l^H \R_{H_l} \F_l \!+ \!\I)^{-1} (\P_l^H\P_l)\},
\label{eq:objective_MSE(F)_reformulated}
\end{equation}
where
$\mathbf{R}_{H_l}
\!=\! \frac{1}{n}\, \mathbf{H}_{l,l}^H \mathbf{R}_{\mathbf{n}_l}^{-1} \mathbf{H}_{l,l}
\in \mathbb{C}^{K N_{T_l} \times K N_{T_l}}$
denotes the effective channel covariance matrix, capturing the impact of the MUIN term. Consequently, $\mathbf{R}_{H_l}$ implicitly depends on the pre-equalizer design of the interfering communication links. The \textit{pre-equalizer only} formulation in $(\ref{eq:objective_MSE(F)_reformulated})$ enables us to equivalently recast $(\ref{eq: non-convex ERM})$   as:
\begin{equation}
\label{eq:non-convex-problem_wrt_F}
\begin{aligned}
\min_{\F_l} \;\;
& {\frac{1}{n}} \operatorname{tr}\!\left\{\left(\F_l^{H} \R_{H_l} \F_l + \I \right)^{-1} \P_l^H \P_l \right\} \\
&\text{s.t.}\quad
 \operatorname{tr}\!\left\{\F_l \F_l^H \right\} \le K P_{\max},
\end{aligned}
\end{equation}
which is still a non-convex optimization problem, but formulated only w.r.t. the pre-equalizer $\F_l$. Interestingly, a closed-form solution to (\ref{eq:non-convex-problem_wrt_F}) is attainable in the absence of compression, i.e., when $K N_{T_l} \!=\! d_l/2$. However, in the more general case $K N_{T_l} \! < \! d_l/2$, we must instead rely on an approximate formulation of (\ref{eq:non-convex-problem_wrt_F}), which yields a closed-form expression for the optimal pre-equalizer. To this aim, let us introduce 
\begin{equation}
\R_{H_l} = \V_{h_l}\mathbf{\Lambda}_{h_l}\V_{h_l}^H \;,\quad
\tilde{\P}_l = \tilde{\U}_{p_l}\tilde{\mathbf{\Sigma}}_{p_l}\tilde{\Q}_{p_l}^H
\label{eq:channel_pilot_decomposition},
\end{equation}
where $\tilde{\mathbf{P}}_l$ denotes the best rank-$K N_{T_l}$ approximation of $\mathbf{P}_l$ in (\ref{eq:non-convex-problem_wrt_F}). Specifically, $\tilde{\U}_{p_l} \in \mathbb{C}^{\frac{m_l}{2} \times KN_{T_l}}$ and $\tilde{\Q}_{p_l} \in \mathbb{C}^{ \frac{d_l}{2} \times  KN_{T_l}}$ in (\ref{eq:channel_pilot_decomposition}) are semi-unitary singular vector matrices, 
performing the best rank approximation
over the first $KN_{T_l}$ orthonormal columns, encoding the principal directions of the semantic pilots cross-covariance. Clearly, if $K N_{T_l} \!=\! d_l/2$, we have $\tilde{\mathbf{P}}_l=\P$.
Now, leveraging the decompositions in (\ref{eq:channel_pilot_decomposition}), we can design $\F_l \in \mathbb{C}^{KN_{T_l} \times \frac{d_l}{2}}$ as:
\begin{align}
    \F_l = \V_{h_l} \mathbf{\Phi}_l \tilde{\Q}^H_{p_l} 
    \label{eq:F_definition}
\end{align}
where $\boldsymbol{\Phi}_l \in \mathbb{R}^{K N_{T_l} \times K N_{T_l}}$ is a diagonal matrix that specifies the power allocation across the $l$-th transmitter's antennas. Then, using (\ref{eq:F_definition}) and substituting $\P$ with $\tilde{\P}$, the objective of \eqref{eq:non-convex-problem_wrt_F} can be approximated as: 
\begin{equation}
\mathrm{MSE}_l(\boldsymbol{\Phi}_l)
\!\approx\! {\frac{1}{n}}\! \operatorname{tr}\!\left\{\!(\mathbf I + \tilde{\mathbf Q}_{p_l}\,|\mathbf{\Phi}_l|^2\,\boldsymbol{\Lambda}_{h_l}\,\tilde{\mathbf Q}_{p_l}^H)^{-1}
\big(\tilde{\mathbf Q}_{p_l}\,\tilde{\boldsymbol{\Sigma}}^2_{p_l}\,\tilde{\mathbf Q}_{p_l}^H\big)\!\right\}\!
\label{eq:mse_Phi_diagonal}
\end{equation}
where $|\mathbf{\Phi}_l|^2 = \mathbf{\Phi}\mathbf{\Phi}^H $. Finally, invoking the matrix inversion lemma and the cyclic invariance of the trace, $(\ref{eq:mse_Phi_diagonal})$ admits a full diagonal reformulation in $\mathbf{\Phi}_l$, leading to the equivalent power allocation problem:
\begin{align}
\max_{ \{ \mathbf{\Phi}_l \}_{l=1}^L} & \quad {\frac{1}{n}} 
\operatorname{tr} \Bigl\{  \Bigl[  (|\mathbf{\Phi}_l|^{2}\boldsymbol{\Lambda}_{h_l} )^{-1} + \I \Bigr]^{-1} \tilde{\mathbf{\Sigma}}^2_{p_l}\Bigr\} 
\notag \\
    \text{s.t.} & \quad \text{tr}\{ \mathbf{\Phi}_l \mathbf{\Phi}_l^H\} \leq KP_{\max}
\label{eq:diagonal_Phi_optimization_problem}
\end{align}
Leveraging this procedure, we obtain a pre-equalizer–only formulation in which all involved matrices are diagonal. In this setting, both the spatial and semantic modes of the logic and physical channels are jointly exploited to determine the optimal transmission directions for semantic messages over the MIMO interference channel. The per-mode gains are given by the power allocation vector $\operatorname{diag}(\boldsymbol{\Phi}_l)$, which serves as the optimization variable.
The resulting scalarized problem of $(\ref{eq:diagonal_Phi_optimization_problem})$ boils down to the maximization of a concave, non-decreasing function $p_l(\cdot)$, by introducing $\boldsymbol{\varphi}_l = \text{diag}(\sqrt{\boldsymbol{\Phi}_{l}})\!=\![\varphi_{l,1},\ldots,\varphi_{l,KN_{T_l}}]^T$, $\text{diag}(\boldsymbol{\Lambda}_{h_l}) \!= \! [\lambda_{l,1},\ldots,\lambda_{l,K N_{T_l}}]^T$, $\text{diag}(\tilde{\boldsymbol{\Sigma}}^2_{p_l}) \!= \! [\tilde{\sigma}_{l,1},\ldots,\tilde{\sigma}_{l,K N_{T_l}}]^T$, for all $l \in L$, thus obtaining: 
\begin{equation}
\begin{aligned}
\boldsymbol{\varphi}^{*}_l
= \arg\max_{\boldsymbol{\varphi}_l}\; p_l(\boldsymbol{\varphi}_l)
= {\frac{1}{n}} \sum_{m=1}^{KN_{T_l}} 
\frac{\bigl(\varphi_{l,m}\lambda_{l,m}\bigr)\tilde{\sigma}^{2}_{l,m}}
{\varphi_{l,m}\lambda_{l,m}+1}
\quad\\
\text{s.t. }\; \varphi_{l,m}\ge 0\;\; \hbox{for all $m$},\quad
\boldsymbol{1}^T \boldsymbol{\varphi}_{l}\le K\,P_{\max}.
\end{aligned}
\label{eq:power-opt}
\end{equation}
Solving (\ref{eq:power-opt}) and hinging on the structured solution in (\ref{eq:F_definition}) and (\ref{eq:Wiener_filter}) for the (pre-)equalizers, semantic channel equalization boils down to a scalar power allocation problem, admitting a low complexity closed-form solution. Indeed, solving the Karush–Kuhn–Tucker (KKT) conditions for (\ref{eq:power-opt}), we derive the optimal scalar power allocation as \cite{boyd2004convex}:
\begin{equation}
  \varphi_{l,m}^{\ast}
  \;=\; \left[\frac{\tilde{\sigma}_{l,m}\,\sqrt{\frac{\lambda_{l,m}}{n\mu_l}} - 1}{\lambda_{l,m}}  \right]_+  \qquad   1\leq m\leq KN_{T_l}, 
\label{eq:closed form solution phi}
\end{equation}
for all $l \in L$, where $[x]_+=\max(0,x)$, and the associated complementarity slackness multiplier $\mu_l \in \mathbb{R}$ can be found via the bisection method imposing the constraint $\boldsymbol{1}^T \boldsymbol{\varphi}_{l}\le K\,P_{\max}$ in (\ref{eq:power-opt}). Interestingly, $p_l(\boldsymbol{\varphi}_l)$ encapsulates the MUIN information within the eigenvalues of $\mathbf{R}_{H_l}$ (cf. (\ref{eq:channel_pilot_decomposition})), and therefore depends directly on the power allocation strategy adopted by the coexisting devices. This formulation allows us to design each $l$-th communication link as a strategic player, acting selfishly to optimize its power allocation vector, assuming the alignment strategies of interfering links remain fixed and denoted as $\boldsymbol{\varphi}_{-l} \triangleq \bigl(\boldsymbol{\varphi}_{1},\, \ldots,\, \boldsymbol{\varphi}_{l-1},\, \boldsymbol{\varphi}_{l+1},\, \ldots,\, \boldsymbol{\varphi}_{L}\bigr)$. Note that each player need only its local estimation of the MUIN covariance in (\ref{eq:MUI_covariance}) (that do not depends on transmitted symbols, due to A0) to compute its locally optimal strategic action. Building on this principle, in the sequel, we introduce a distributed non-cooperative game-theoretic formulation that enables semantic alignment and coexistence of multiple AI-native devices over interference channels.



\subsection{Distributed Games for Semantic Equalization}
The context described so far clearly establishes, within a game-theoretic framework, that interactions among players occur through their objective functions. By modeling each $l$-th communication link as a player whose decision variable depends on the strategies of its rivals, we can formulate the problem as a Nash Equilibrium Problem (NEP). Formally, we consider $L$ communication links modeled as strategic players. Each player $l$ controls its own decision variable $\boldsymbol{\varphi}_l \in \mathbb{R}^{K N_{T_l}}$, constrained to a local strategy set $\mathcal{Q}_l \subseteq \mathbb{R}^{K N_{T_l}}$. Since no coupling constraints are present among players, the global feasible set is the Cartesian product $\mathcal{Q} \triangleq \mathcal{Q}_1 \times \dots \times \mathcal{Q}_L \subseteq \mathbb{R}^{K \sum_{l=1}^{L} N_{T_l}}$. Accordingly, the aggregated decision variable is denoted by $\boldsymbol{\varphi} \triangleq (\boldsymbol{\varphi}_1, \dots, \boldsymbol{\varphi}_L) \in \mathcal{Q}$.
 \begin{algorithm}[ht]
    \caption{\small Distributed Games for Semantic Equalization}
\label{alg:zs-sce-2e}
\DontPrintSemicolon
\KwIn{$ \F_l^{(0)}\sim \mathcal{CN}(0,1)$ s.t. $||\F_l||^2=KP_{\max}$, $\forall l$;}
\KwOut{Optimal semantic equalizers $\F_l$ and $\G_l$, $\forall l$;}
\For{$t \gets 1$ in \text{game iterations}}{
\If{$\F_l^{(t)}$ \text{satisfies 
termination criterion}}
{\text{STOP}}
\Else{
  Compute $\G_l^{(t)}$ by (\ref{eq:Wiener_filter}) ;\;
  \If{$\mathcal{B}_l=\text{Gauss-Seidel}$}{
  Compute (\ref{eq:closed form solution phi}) with the update rule (\ref{eq:gauss_seidel_scheme});}
   \ElseIf{$\mathcal{B}_l=\text{Jacobi}$}{
  Compute (\ref{eq:closed form solution phi}) with the update rule (\ref{eq:jacobi_scheme});}
}
Compute $\boldsymbol{\varphi}_l^{t+1}$ by (\ref{eq:best-response step size}), and
set $\F_l^{(t+1)}$ as (\ref{eq:F_definition}). \;
}
\Return{$\Phi^{\star}_l$}\;
\label{algo: potential game for SCE}
\end{algorithm}
Each $l$-th player aims to choose a profile strategy that maximizes its generic payoff function $p_l$:
\begin{equation}
\begin{aligned}
\max_{\boldsymbol{\varphi}_l \in \mathcal{Q}_l}\;\; & p_l\!\left(\boldsymbol{\varphi}_l,\boldsymbol{\varphi}_{-l}\right). 
\end{aligned}
\label{eq:NE-problem}
\end{equation}
An aggregate strategy profile $\boldsymbol{\varphi}^{\star}$ is said to be a NE if no player can improve its objective by unilaterally deviating from its own equilibrium strategy $\boldsymbol{\varphi}_l^{\star}$:
\begin{equation}
    p_l(\boldsymbol{\varphi}_l^{\star}, \boldsymbol{\varphi}_{-l}^{\star}) \geq p_l(\boldsymbol{\varphi}_l, \boldsymbol{\varphi}_{-l}^{\star}),  \quad \forall \boldsymbol{\varphi}_l \in \mathcal{Q}_l.
    \label{eq: NE_condition}
\end{equation}
Generally, the achievability and convergence toward a NE is not guaranteed. Following \textit{Rosen's theorem} \cite{rosen1965existence}, a game $\mathcal{G}: \langle L,\mathcal{Q}, \{u_l\}_{l \in L} \rangle$ admits at least one (pure) NE if:
\begin{itemize}
    \item[(i)] \textit{for every player, the payoff function is continuously differentiable and concave in $(\boldsymbol{\varphi}_l, \boldsymbol{\varphi}_{-l})$ given the strategies of other players $\boldsymbol{\varphi}_{-l}$ };
    \item[(ii)] \textit{each $l$-th nonempty feasible set $\mathcal{Q}_l$ is compact and convex.}
\end{itemize}
Straightforwardly, recalling problem~(\ref{eq:power-opt}), each player’s feasible set $\mathcal{Q}_l$ is convex and compact. 

Moreover, the utility $p_l(\boldsymbol{\varphi}_l, \boldsymbol{\varphi}_{-l})$ is concave in $\boldsymbol{\varphi}_l$ for fixed rivals’ strategies. Therefore, the game admits at least one (pure) Nash Equilibrium. In a NE game, each communication link (player) selfishly maximizes its own payoff function~(\ref{eq:power-opt}), resulting in a distributed best-response dynamics. That is, given the allocation of other users, every player iteratively solves its own optimization subproblem, whose optimal solution is given in closed form by~(\ref{eq:closed form solution phi}). Two different approaches can be followed in the choice of the iterative algorithms to be performed by the players, namely Gauss–Seidel and Jacobi-based schemes \cite{bertsekas2015parallel}. Letting $B_l(\boldsymbol{\varphi}_{-l})$ be the best response mapping in (\ref{eq:closed form solution phi}) for user $l$, considering the Gauss-Seidel algorithm, the players update their strategy sequentially over time:
\begin{equation}
\hat{\boldsymbol{\varphi}}_{l}(
\boldsymbol{\varphi}_{-l})
 \!= \! B_{l}\!\left(
 \boldsymbol{\varphi}_{1}^{\,t+1},\ldots,\boldsymbol{\varphi}_{l-1}^{\,t+1},
 \boldsymbol{\varphi}_{l},
 \boldsymbol{\varphi}_{l+1}^{\,t},\ldots,\boldsymbol{\varphi}_{L}^{\,t}
 \right).
\label{eq:gauss_seidel_scheme}
\end{equation}
Instead, in the Jacobi procedure, all the players optimize their own strategies in a simultaneous manner at time $t$:
\begin{equation}
\hat{\boldsymbol{\varphi}}_{l}(
\boldsymbol{\varphi}_{-l})
 \! = \! B_{l}\!\left(
 \boldsymbol{\varphi}_{1}^{\,t},\ldots,\boldsymbol{\varphi}_{l-1}^{\,t},
 \boldsymbol{\varphi}_{l},
 \boldsymbol{\varphi}_{l+1}^{\,t},\ldots,\boldsymbol{\varphi}_{L}^{\,t}
 \right),
\label{eq:jacobi_scheme}
\end{equation}
for all $l=1,\ldots,L$. In the sequel, we assume the sequence $\{\boldsymbol{\varphi}_l^t\}_{l=1}^L$ admits a limit point; a fact that will be validated by extensive numerical simulations. To better control convergence behavior, we also introduce a step-size procedure given by:
\begin{equation}
    \boldsymbol{\varphi}_{l}^{t+1}= \boldsymbol{\varphi}_{l}^{t} + \gamma^{t} (\hat{\boldsymbol{\varphi}}_{l}(\boldsymbol{\varphi}_{-l}) - \boldsymbol{\varphi}_{l}^{t})
    \label{eq:best-response step size}
\end{equation}
with $\gamma \in (0,1], \gamma^t\xrightarrow{}0$ and $\sum_{t=1}^\infty \gamma^t\!=\!+\infty$. 
In our experiments, we found it particularly effective to update the step size as: $\gamma^{t+1}\!=\! \gamma^{t}(1-\epsilon\gamma^{t})$ with sufficiently small $\epsilon\in (0,1]$.
Our algorithmic procedure to design distributed games for latent space alignment is summarized in Algorithm (\ref{algo: potential game for SCE}). In the sequel, we assess the performance of the proposed methods via numerical simulations.

\begin{figure}[t]
  \centering
\includegraphics[trim=0.30cm 0.41cm 0.15cm 0.25cm,clip,width=1.0\columnwidth]{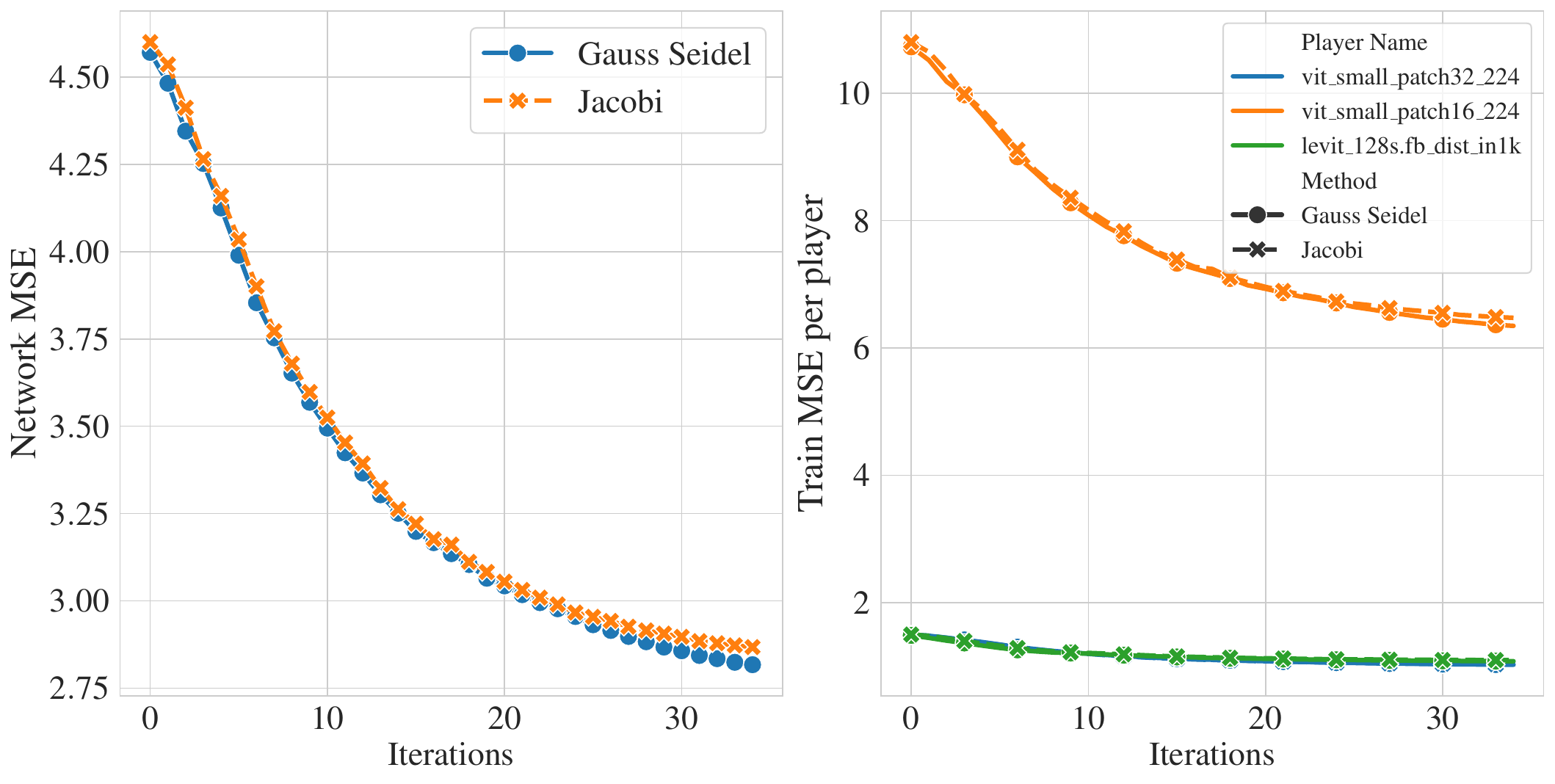}
  \caption{System BR-iterative behavior with $N_{T_l},N_{R_l}\!=\!8$, $K\!=\!10$, $\alpha\!=\!3$. Network (left) and per player (right) $\text{MSE}_l(\boldsymbol{\Phi}_l)$ over game iterations:\textit{ vit\_small\_patch16} 
  is the tx deployed in the middle of the system, suffering the most MUI.}
  \label{fig::mse_game}
\end{figure}

\section{Numerical Results}
\begin{figure}[t]
  \centering
\includegraphics[trim=0.30cm 0.41cm 0.05cm 0.25cm,clip,width=1.0\columnwidth]{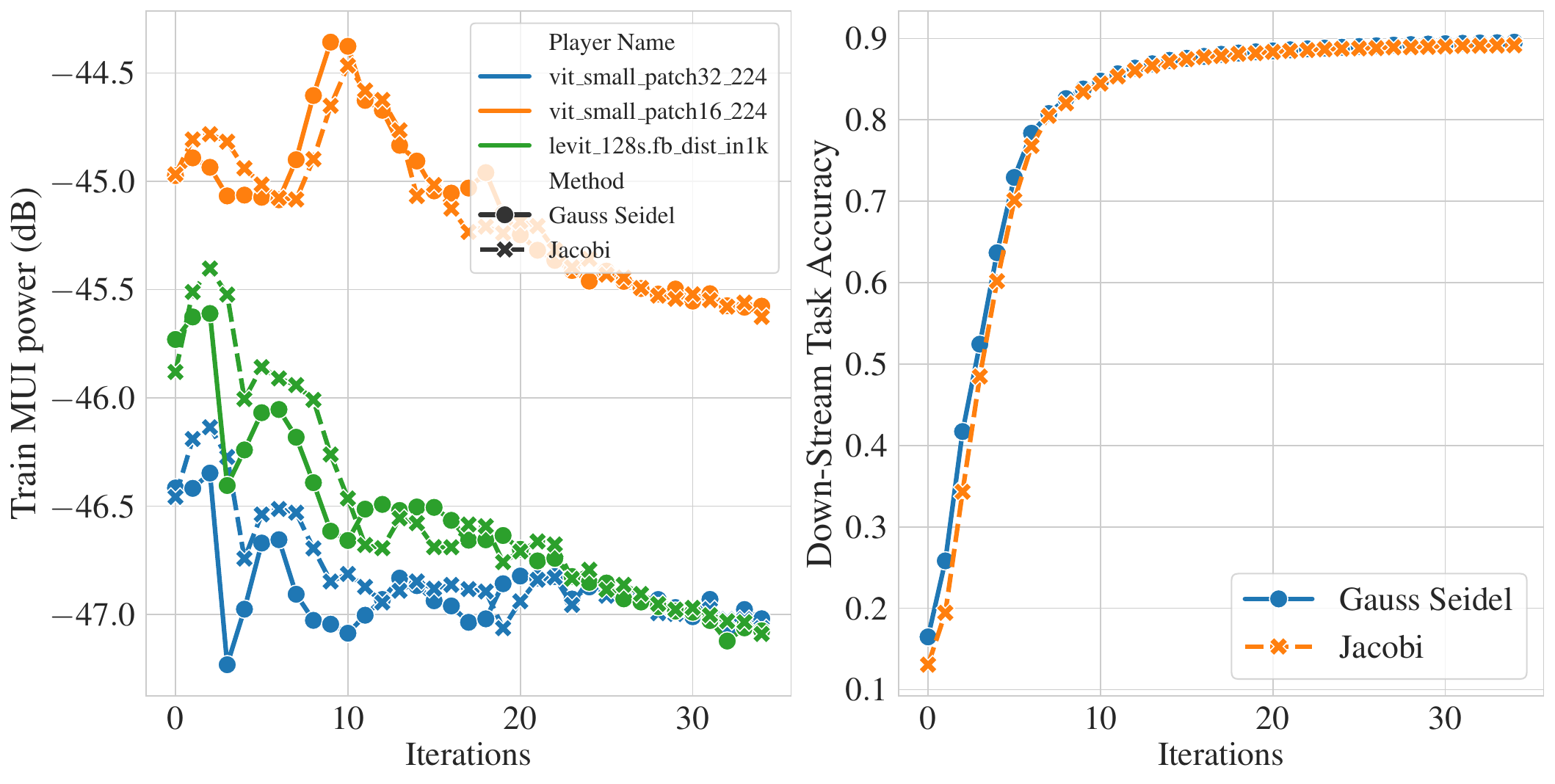}
  \caption{System BR-iterative behavior with $N_{T_l},N_{R_l}\!=\!8$, $K\!=\!10$, $\alpha\!=\!3$. MUI power in dB (left) and network down-stream task accuracy (right) over game iterations.}
  \label{fig::mui_game}
\end{figure}

\begin{figure}[t]
    \centering    \includegraphics[width=\columnwidth]{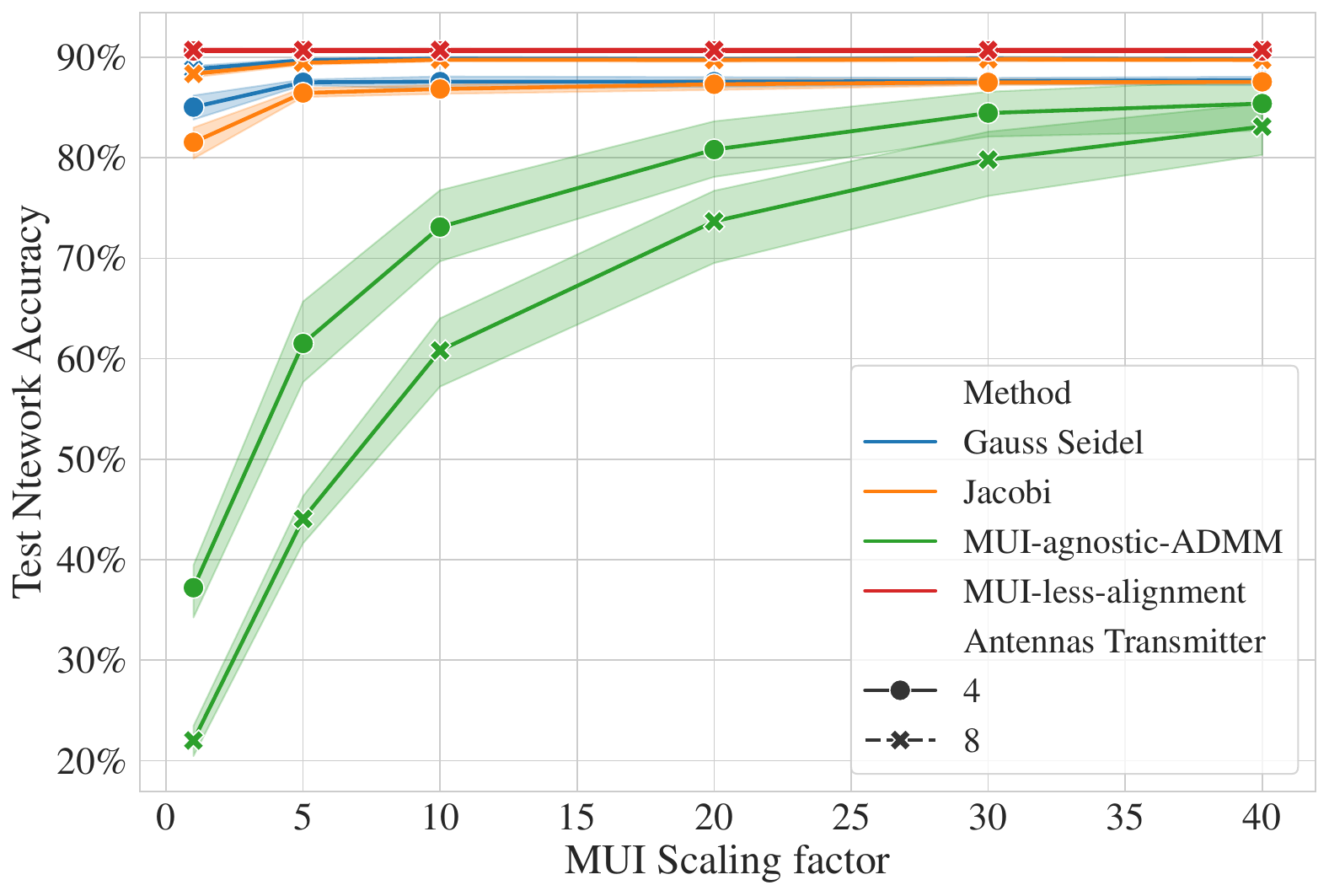}
    \caption{\small Network Accuracy vs. $\alpha$ (MUI scaling factor), with $N_{T_l} = N_{R_l} \!=\! 4,8$ and wireless channel usage $K\!=\!10$.}
    \label{fig::scaling_fact}
\end{figure}

This section provides numerical results
to assess the performance of the proposed distributed games for latent space alignment in a MIMO downlink communication system, considering Rician flat-fading channels with rice factor $\!=\!1.5$, path loss exponent $\!=\!2.5$, and $N_{T_l}\!=\!N_{R_l}\!=\!8$.
Semantic communication is implemented by transmitting latent representations to perform an image classification task. We consider the CIFAR-10 dataset, comprising $32\times32$ color images distributed across 10 classes. Among them, 42500 images were used for training and 10000 for testing, with classification across 10 labels.
Considered latent representations are produced by the backbone of pre-trained models chosen from the \href{https://pypi.org/project/timm/}{\textit{timm}} Python library. 
Precisely, we deploy three MIMO point-to-point links operating at carrier frequency equal to $3.5 \text{ GHz}$, where each intended transmitter–receiver separation is $30$m and the distance between adjacent transmitters varies from $30$m to $1.2$km.
To modulate the strength of MUI term, let us define a \textit{MUI scaling factor} $\alpha \!=\! \frac{d(T_i,T_j)}{d(T_i,R_i)}$  where the numerator represents the distance between the $i$-th transmitter and $j$-th interferer and the denominator the distance between the intended $i$-th transmitter-receiver. The scaling factor parametrizes the strength of the MUI term, such that as it increases, the MUI term will be attenuated.
 Transmitters operate at tx-power $P_{\max}\!=\!1.0$ W per channel use, with their internal logic given by the DNNs vit\_small\_patch32, vit\_small\_patch16, levit\_128s\.fb; receivers interpret latent codes using vit\_base\_patch16, vit\_tiny\_patch16, vit\_base\_patch32\_clip. All displayed results are averaged across channel realizations given by seeds \{27-42-100-123-144-200\} with confidence bands that highlight the variability across different channel realizations. For the sake of simplicity, we choose internal logic operating in transmission holding latent spaces of equal dimensions, to define a common compression factor $\xi$. In Fig. \ref{fig::mse_game}, \ref{fig::mui_game} we illustrate mean squared error, MUI term and network accuracy of the proposed game theoretic approach over iterations, highlighting convergence to limit point of the best-response generated sequence.
\begin{figure}[t]
    \centering    \includegraphics[trim=0.30cm 0.32cm 0.15cm 0.25cm,clip,width=\columnwidth]{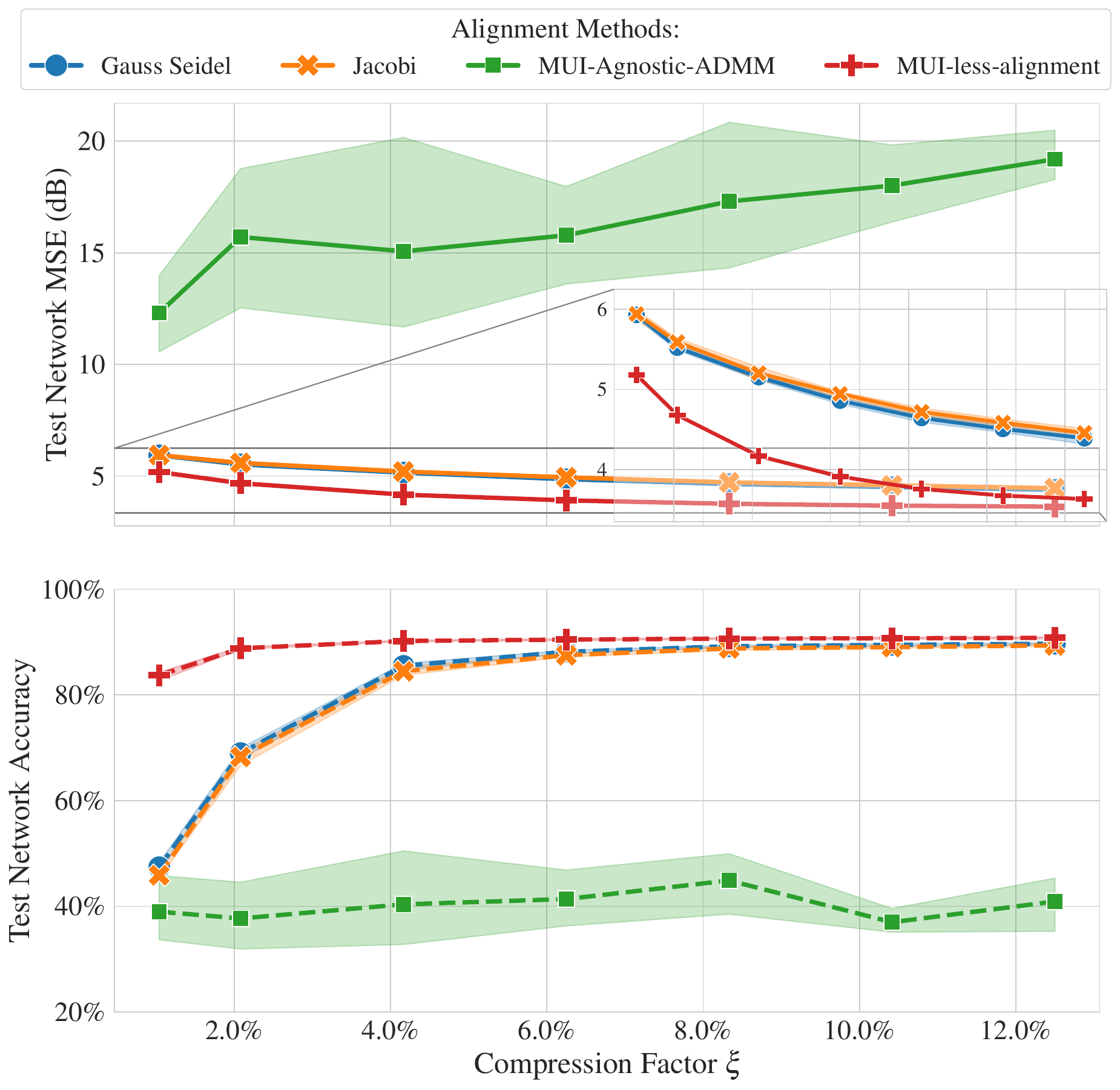}
    \caption{\small Accuracy-MSE vs. $\xi$, with $N_{T_l}\! =\! N_{R_l} \!=\! 8$ and $\alpha\!=\!3$.}
\label{fig::compression_fact}
\end{figure}
Interestingly, in Fig.\ref{fig::mui_game} the link with higher MUI is the one deployed in the middle of the system, therefore suffering the highest interference.
It represents the struggler of the network, in which players design their pre-coding function in an attempt to mitigate the interference term in relation to it, as it is shown by the decreasing trend of the per player MUI. 
We compare the proposed game-theoretic method
to two benchmarks, respectively, the case of pre-equalizer only alignment formulation (MUI-less-Alignment), ideally evaluated without MUI, therefore considering $\R_n \! = \! \R_v$, and an Alternating direction method of multipliers (MUI-agnostic ADMM) driven alignment, as implemented in \cite{pandolfo2025latent}, in which MIMO Transceivers are derived optimizing in an alternating fashion, leveraging the block-convex structure in (\ref{eq: non-convex ERM}).
\noindent Fig. \ref{fig::compression_fact} shows the MSE (dB) 
together with downstream-task accuracy.  The results highlight that the proposed game-theoretic framework substantially mitigates multi-user interference compared to MUI-agnostic aligners, as the channel usage increases, it is able to get closer to the MUI-less Alignment benchmark.
Finally, Fig. \ref{fig::scaling_fact} highlights the interference mitigation capabilities with different antenna configurations, with network accuracy as a function of $\alpha$ (i.e. the MUI scaling factor): 
the proposed approach can get significantly close to the MUI-less case, showing in the meantime a significant gap with an MUI-agnostic approach, which is not able to orthogonalize the MUI term as coexisting links get closer (lower $\alpha$), leading to catastrophic task execution. Remarkably, MUI-agnostic aligners perform even worst, especially as interferes links get closer, increasing the number of transmitter antennas, as shown in Fig. \ref{fig::scaling_fact}, due to the higher perceived interference at the intended receiver. This behavior further enforce robustness of the proposed game-theoretic approach, that effectively promotes interference mitigation, while enabling information compression and alignment.

\section{Conclusions 
}
In this paper, we proposed a unified framework for multi-user semantic communication that jointly addresses latent-space misalignment and semantic coexistence under interference, enabled through end-to-end optimized MIMO semantic transceivers. Our results demonstrated that semantic noise, induced by misaligned latent representations across heterogeneous semantic transceivers, can be effectively mitigated when semantic alignment is jointly optimized with transmission strategies. By formulating the interaction among users as a distributed non-cooperative game, we derived a closed-form solution that enables coordinated semantic equalization, even in interference-limited settings. 
Numerical evaluations validated the proposed goal-oriented communication paradigm, revealing key trade-offs among information compression, interference mitigation, semantic alignment, and downstream task performance. Future work will investigate convergence guarantees for iterative best-response dynamics and develop scalable mechanisms for semantic coexistence, including support for primary and secondary semantic communication users.

\bibliographystyle{IEEEtran}
\bibliography{ref}
\end{document}